# Rastro-DM: data mining with a trail

## A methodology for documenting data mining projects and its application in the construction of a text classifier of documents associated with damages to the public treasury

**MARCUS VINÍCIUS BORELA DE CASTRO 1**

Auditor at the Court of Accounts of Brazil (TCU) since 1996. Bachelor's degree in Informatics from the Federal University of Viçosa (1990), and specialist in IT Governance from the University of Brasilia (2012), and in Data Analysis from the Brazilian Federal Court of Account's Serzedello Corrêa Institute (2019).

**REMIS BALANIUK**

Auditor at the Court of Accounts of Brazil since 1989. Bachelor's degree in Computer Science from the University of Brasilia (1986), Master's degree in Computer Science from UFRGS (1989), doctoral degree in Informatics from the Institut National Polytechnique of Grenoble (1996), and postdoctoral research in Computer Science at Stanford University (2002) and at the Institut National pour la Recherche en Informatique et Automatique (2000). Visiting researcher at the University of Oxford (2020).

## ABSTRACT

This paper proposes a methodology for documenting data mining (DM) projects, Rastro-DM (Trail Data Mining), with a focus not on the model that is generated, but on the processes behind its construction, in order to leave a trail (Rastro in Portuguese) of planned actions, training completed, results obtained, and lessons learned. The proposed practices are complementary to structuring methodologies of DM, such as CRISP-DM, which establish a methodological and paradigmatic framework for the DM process. The application of best practices and their benefits is illustrated in a project called "Cladop" that was created for the classification of PDF documents associated with the investigative process of damages to the Brazilian Federal Public Treasury. Building the Rastro-DM kit in the context of a project is a small step that can lead to an institutional leap to be achieved by sharing and using the trail across the enterprise.

**Keywords:** Data mining; Data analysis; Data science; Machine learning; Organizational knowledge; Methodology; Best Practices; Data analysis in Government; Textual classification; Documentation; Documentation of data mining projects.

---

1   This is a concise version of the conclusion work resulting from the postgraduate lato sensu work in Data Analysis indicated in the references (Castro, 2019).

## 1. INTRODUCTION

Data are changing everything and the capacity of manipulating data and understanding Data Science is becoming increasingly critical for current and future discoveries and innovations (BERMAN et al, 2018).

Data mining (DM) projects are challenging not only because of the complex and exploratory process used on them, but also because they are generally innovative, unique and very often developed by individuals or small teams.

These are innovative projects, either because they use techniques and algorithms that may not be consolidated in the Organization or in academic research, or because they require the construction of models that simulate cognitive processes and natural intelligence by machines.

These projects are usually unique. The particularities of each context, the data implicated, and the quality requirements, prevent or make difficult the reuse of codes from other projects.

They are complex projects because the techniques used typically have a conceptualization difficult to understand and require interdisciplinary knowledge of areas such as computer science, mathematics, and statistics. Furthermore, it requires an understanding of the business for which the solution is designed.

They are exploratory processes, because the data mining activity can be defined as the process of exploring combined data with different techniques, extracting or helping to demonstrate patterns and assisting in the discovery of knowledge.

Additionally, in the case of organizations with a low maturity level in DM, these projects are the responsibility of small teams or even of individual persons. In this case, the sharing of knowledge and practices adopted in the projects becomes even more difficult.

Unfortunately, these complex works, of exploratory nature, that are innovative, unique, and, typically, individualized, do not leave a trail of what they delivered. In the end, we have the implemented solution. There may even be documentation relative to the product created. However, there is none regarding the process adopted, the choices made, and the techniques used in the different activities of the project. We are left with concerns in case it is necessary to reconstruct the model without the presence of its creator. In addition, the person responsible for it becomes the parent of the product, since only this person knows it and can maintain it.

One intriguing debate in the management of general knowledge is how to collect, gather or make explicit the experience of project development for it to be applicable for others (DINGSØYR et al, 2001). Because the memory of an organization cannot be based uniquely in the memory of its individuals (STATA, 1980).

In face of the aforementioned, a challenging topic arises: How to systematize the documentation of the tasks of a data mining project in a manner that can leverage its auditing capability and the sharing of knowledge?

In order to answer the question posed, the objective of this work is to propose a methodology as a set of good practices for the semi-automated tracking of the activities of a DM project, in order to provide a trail of the choices made, of the data processing carried out, and of the results obtained, not focusing on the product generated, but instead, focusing on the process underlying its creation. These are good practices that can be combined with the corporate DM methodology being used in the organization. The production of trails in DM projects accelerates the learning curve and the creation of an organizational culture centered on the use of data analysis.

There are three specific objectives: contextualizing the theoretical framework relative to the methodologies and documentation in data mining projects as well as the potential impact of the experiences acquired in the projects for an organization; proposing the Rastro-DM (Trail Data Mining) methodology with the description of its activities; and the illustration of its application in a project for the classification of PDF documents associated to damages to the Brazilian Federal Public Treasury, named "Cladop", demonstrating its feasibility and the benefits of its use. The specific objectives will be discussed in the three sections of the body.

## 2. THEORETICAL FRAMEWORK

### 2.1 DATA MINING METHODOLOGIES

According to BERMAN et al (2018), the Science of Data is concentrated in the processes of extraction of knowledge or insights, originating from structured or non-structured data. This process of discovery of data knowledge (KDD - Knowledge-Discovery in Databases), according to BECKER and GHEDINI (2005), is complex and popularly called DM - Data Mining. In the scope of this work, we will refer to these processes indistinctly as KDD of DM.

CHAPMAN et al (2000) specify some objectives for DM translated into groups of problems to be discussed: description and summarization of data, segmentation (clustering), description of concepts, classification, prediction (regression), and analysis of dependence.

WIRTH and HIPP (2000) confirm that DM is a complex process and associate the success of a project to an adequate combination of good hardware, qualified analysts, and use of a solid methodology and of effective project management. With regards to the methodology, the same authors state that DM requires a standard approach that assists in transforming business problems into data mining tasks, that suggests appropriate transformations of data and techniques to be employed, and that provides the means to evaluate the effectiveness of the results and document the experience. Moreover, they emphasize that the use of a methodology in planning and presenting reports inspires confidence to users and to sponsors.

KURGAN and MUSILEK (2006) state that methodologies provide a better comprehension and understanding of the process. Furthermore, they promote economy of time and costs, with the establishment of an itinerary to be followed for the planning and the execution of the projects. However, they mention that a great number of projects follow on their own methodologies.

An example of a methodology used in DM is the CRISP-DM (Cross-Industry Standard Process for Data Mining). It is considered a standard and one of the factors of its success is its neutrality in terms of industry, hardware, and applications (MARISCAL *et al*, 2010).

CHAPMAN *et al* (2000) corroborate that the CRISP-DM methodology is a model of hierarchical process, consisting in combined tasks described in four levels of abstraction (from generalized to more specific levels): phase, generic task, specialized task, and process instance. They state that the sequence of the phases is not restrictive. That, from a practical point of view, many tasks can be executed in a different order, and, frequently, it is necessary to return repeatedly to the preceding tasks and redo specific actions. They emphasize that representing every possible sequence of the means of the data mining process would require an excessively complex model of processing. Therefore, they do not do that. They state that a phase is never completely concluded before the following phase begins. Figure 1 displays the iteration between phases of the CRISP-DM methodology.

Figure 1 – General view of the CRISP-DM methodology.

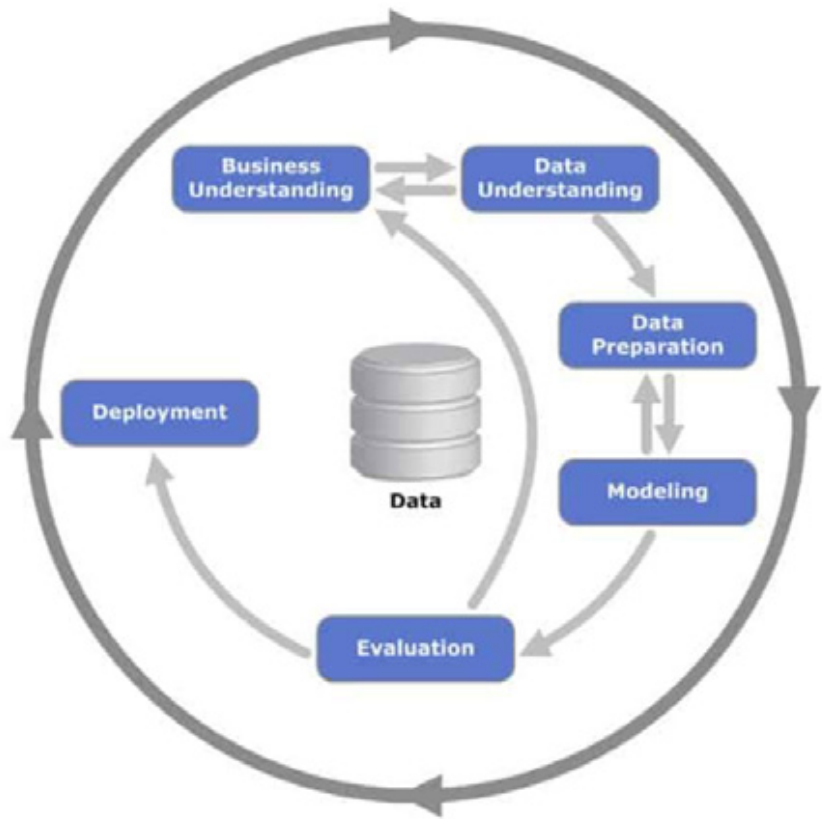

Source: CHAPMAN et al (2000).

According to MARBÁN et al (2007), the CRISP-DM methodology does not encompass many tasks related to management, organization, and project quality. At least, not in a manner demanded by the increasing complexity of current DM projects that include not only a great volume of data, but also the management and organization of large interdisciplinary teams. These authors group the tasks of a DM process, relative to the construction of a model, into three stages: pre-development, development, and post-development. They state that all the methodologies used in DM are concentrated in the development stage, which corresponds to the collection and analysis of the available data for the project, the creation of new data based on the available data, the adaptation for DM algorithms, and the creation of models.

According to CHOLLET (2017), the development of a project is concentrated in the experimentation of a model: it begins with an idea that is expressed as an experiment, in the attempt to validate or invalidate the idea. Subsequently, this experience is executed, and the information generated is processed. That will inspire the following idea. The more iterations in this repetitive circle, more refined and powerful the ideas will become. He emphasizes the importance of obtaining the maximum amount of information from the experimentation, including the performance of the models.

The experimentation, that is, the application of mathematical algorithms to the data for the extraction of patterns, is called training by NGUYEN (2018). In order to simplify, in the context of this work, the terms training and model will be used to represent the experimentation and the product resulting from the patterns detected in the data.

GREFF et al (2017) confirm the substantial number of computational experiments with several different configurations of hyperparameters and warn against the practical challenge of documentation. Because of the pressure regarding deadlines and the inherently unpredictable nature of a DM project, they verified that there is little incentive for the construction of robust infrastructures and, as a result, the coding will frequently evolve rapidly, which compromises, among other things, the documentation of the project.

BECKER and GHEDINI (2005) emphasize that the structure of a DM process is highly dependent on the methodology adopted, on the abilities, experience and approach of the person responsible for the process, as well as on the resources available in the corporation. Additionally, they confirm the high interactivity of the processes. They emphasize that, even though the conceptual structure of the process may suggest an order between the phases, in practice, the analysts go from one phase to almost any other random phase at any moment. This is also because several problems related to the preceding phases (for example, preparation of data) can only be detected a long time later, when the patterns and the models are evaluated. The same authors also state that generally the projects are developed in a nonstructural manner, ad hoc: after the initial analysis of the data, it is decided to test a specific technique, from which the results may suggest the restructuration of data and the execution of new types of analysis; and so on and so forth. Additionally, over time, it is difficult to remember which training was carried out, which sets of data and which hyperparameters were used, and, more important, the results derived from the combined data. Moreover, according to the authors, this situation leads to recurrent execution of training. They advise that the situation is even worse when taking into consideration long term projects involving several people. They have verified that, regardless of the diversity of knowledge of different people, techniques and instruments, the majority of the DM projects face, in practice, the same difficulties: the wastefulness of redoing the same work and the management of resources and results. The authors state that the documentation of the history of the tasks in face of the iteration and interactivity of the process is an open problem in the management of DM projects.

## 2.2 DOCUMENTATION: A PATH FOR THE GENERATION OF KNOWLEDGE THAT CAN BE SHARED

GHEDINI and BECKER (2000) emphasized that the documentation of tests and of all relevant parts of a project can not only avoid the loss of the knowledge confined in the minds of the people but can also allow its sharing, turning into a rich source of knowledge for future references and corporate reuse. They also emphasize that this activity leads to a better management of efforts, resources, and results of a DM project.

According to PRAKASH et al (2012), information from training and implementation code (and the history of change) contain an abundance of information regarding the state, progress and evolution of a software project. Additionally, they state that DM is becoming an increasingly important tool to transform these data into information. Similarly, it is expected that the data from DM projects will also be converted into information by the activities of data mining.

WIRTH and HIPP (2000) emphasized that, perhaps, the greatest benefit of having applied a methodology was the documentation generated. They admitted having initially skipped some tasks of documentation and planning because they would take too long and because they were considered unnecessary by specialists such as them. However, they presented the price that they payed due to this action and they regretted it, realizing that every effort is worthwhile. They reported some benefits observed based on the documentation produced: it avoids the wastefulness of effort (for example, in unfruitful paths or with repetitive work); it promotes efficient management and better communication by the team; it allows for the identification of critical points in the process; it promotes better planning for future projects, based on a more accurate perception of how the effort was spent and of the necessary resources; and it promotes the use of experiences documented in other contexts.

BECKER and GHEDINI (2005) also identified the role of the documentation in learning experiences and in the reuse, and state that an immediate benefit of documentation is the effectiveness in management, in planning, and in communication. They verified that the documentation is completely dependent on the project team because the veracity of the registration and the level of detail that will directly impact its usefulness is their responsibility. Regarding resistance, they state that, insofar as the benefits of the activity are perceived, there is incentive to document the process with more detail and in parallel to the execution of the activities. Additionally, that the best results are reached in the long term, when the project team discovers which strategy is more adequate to its work methods, as well as the best ways to obtain advantages from the resources, techniques and flexibility of the model.

Nevertheless, WIRTH and HIPP (2000) advise of the difficulty of proceeding to the documentation in the end, and of trying to reconstruct what had been carried out and its motivations. They emphasize that DM processes are live and, given this circumstance, the documentation shall be flexible and live, and should never be updated subsequently to the end of the project (postmortem). They recommend that the definition of a strategy for documentation must be the starting point, nonetheless, the flexibility for evolution and change must be a premise. They emphasize that to meet the right level of detail for planning and documenting a DM process is difficult and it is part of a long learning process, and it can be influenced by several factors such as the complexity of the project, the duration, and the size of the team.

GREFF et al (2017) reference that one of the difficulties for sharing, collaboration, and reproducibility of the training is the use of particular configurations of the workplace by the teams, when using the different tools necessary to discuss the different aspects of a DM process. They mention, among others, databanks, source code management, automatized tools for optimization of hyperparameters, scripts and tables. The same authors presented Sacred, a Python framework of open code that aims to provide basic infrastructure for the execution of computational experiments irrespective of the methods and libraries utilized. They are concentrated on solving problems such as the management of configurations, documentation, and the reproducibility of the results. For each training, relevant information, such as parameters, package dependencies, host information, source code, and results are automatically captured and stored in a centralized repository, where details about the hyperparameters used and the results obtained can also be consulted.

PRAKASH et al (2012) report that several works are being executed in the development of integrated platforms for Machine Learning (ML) and for Software Engineering based on reusable components, quoting, among the most prominent ones with open codes, WEKA and Rapid Miner.

PUBLIO et al (2018) believe that the vision of canonical and standardized models may lead to a better comprehension of ML data and algorithms employed in DM and may promote the interoperability of experiments, irrespective of the platform or of the solution of workflow adopted.

The W3C Machine Learning Schema Community Group (2017) published a ML Schema ontology that provides a set of categories, properties and restrictions for representing and interchanging information about algorithms of machine learning, data sets, and experiments. According to the group, the ontology can be easily extended and mapped for other more specific ontologies of domain, developed in the field of machine learning and data mining. Figure 2 presents the visualization of the main abstractions of the ML Schema.

Figure 2: Main concepts of the ML Schema.

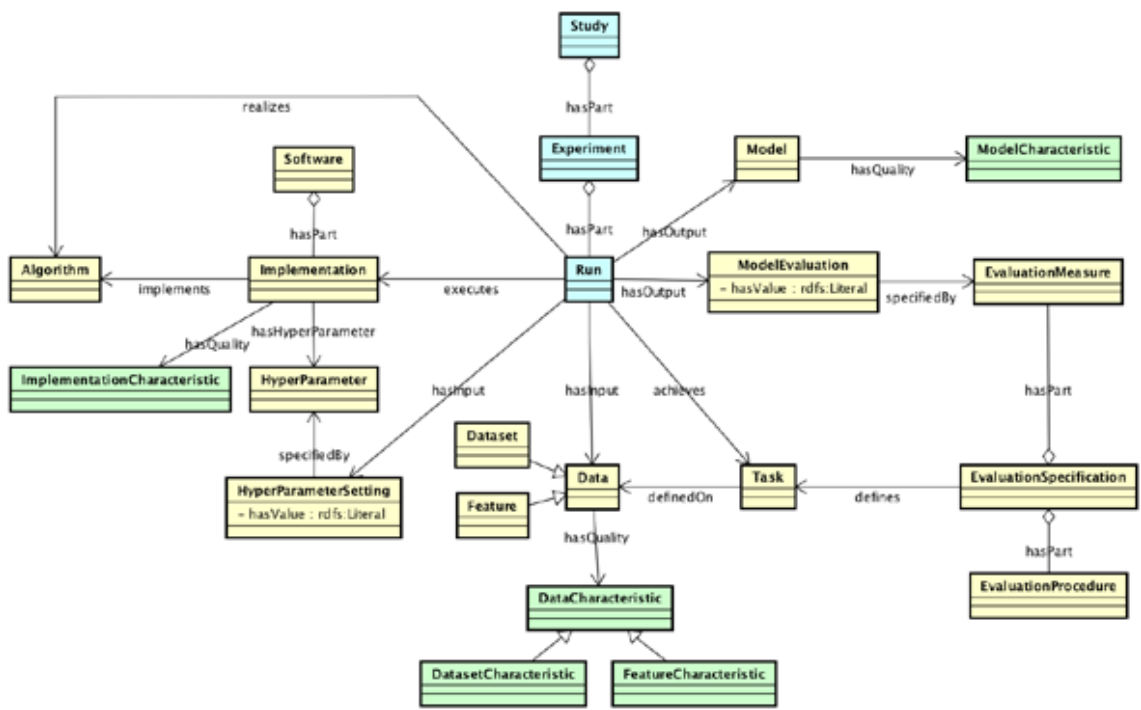

Source: W3C Machine Learning Schema Community Group (2017).

Concisely, the central concept is the training (Run) that, in the context of an experiment (Experiment, Study), produces a model (Model), with its characteristics (ModelCharacteristic), and a set of quality evaluations (ModelEvaluation) that take into consideration standardized metrics (EvaluationMeasure, EvaluationSpecification, EvauationProcedure). The training is the execution of the implementation (Implementation) of an algorithm (Algorithm) in a platform (Software, ImplementationCharacteristic) using specific parameters configuration (HyperParameter, HyperParameterSetting). The training has as its entrance a dataset (Data, Dataset, Feature) that has its own characteristics (DataCharacteristic , DatasetCharacteristic, Featurecharacteristic) and they are utilized for a specific task (Task).

KURGAN and MUSILEK (2006) refer to the possibility of integration and of interoperability of DM models using industrial patterns such as the PMML (Predictive Model Markup Language), which represents a model in an XML scheme (Extensible Markup Language). According to the authors, different tools can be used for the generation, visualization, and analysis of the same model.

## 2.3   KNOWLEDGE IN A DM PROJECT: A POTENTIAL ORGANIZATIONAL LEAP FORWARD

WIRTH and HIPP (2000) emphasize that the success or the failure of a data-mining project is highly dependent upon the person or upon the team, and that successful practices are not necessarily repeated within the entire company.

CHAPMAN et al (2000) emphasize that data-mining projects can benefit from the experiences from previous projects. The lessons learned during the process and deriving from the solution implemented can give rise to new business debates that are frequently more targeted.

BECKER and GHEDINI (2005) exemplify the benefits derived from the sharing of knowledge between projects and state that the experiences from previous projects can be used to establish plans for more realistic DM projects, with more precise estimates of schedules, budgeting, etc. Experience develops into an easier way to favor more realistic resources, because there is a more extensive understanding of how the effort is actually spent. Furthermore, previous experiences can be used to deal with certain categories of problems or techniques. They advocate that the documentation of the execution of projects must be treated as a corporate resource, which can be shared by the team and used as a reference and can be dependent upon corporative policies and standards.

According to BHATT (2001), while the individuals in the organizations interact with each other, they tend to understand and share their different visions regarding similar situations, structuring their communities and sharing effective work techniques and facilitating the integration of a diversified body of knowledge in the organizations. The same author states that organizational knowledge is formed by unique patterns of interactions between technologies, techniques, and persons, which cannot be easily reproduced, because these interactions are unique from a specific organization, modeled by its history and by its culture. The author attributes the sustainability of the competitive advantages of the company in the long term to incentive given for the enlargement of this knowledge by the creation of a stimulating and practical environment (to learn by practice).

DINGSØYR et al (2001) discuss in an indistinct manner knowledge and experience. They recognize that experience in the strict sense of the word is something that resides in human beings and that it cannot be transferred to others (that each one would have to experiment by herself of himself to acquire the experience). Nevertheless, in a less strict definition, they state that experience is information, which is operational, that is to say, it is usable in some situations. They understand that a description of an event that has happened during a project is one item of experience.

In view of the above mentioned in the introduction, an interesting debate in the management of general knowledge is how to collect, gather, or make explicit experiences in projects in order to be workable for others (DINGSØYR et al, 2001).

NGUYEN (2018) recommends the use of KMP - Knowledge Management Process to have knowledge circulating throughout the entire organization in order to ensure that the right knowledge reaches the right person to understand and have knowledge enough to make decisions and adequately implement tasks. The author states that KMP can be used at any level, from the organization as a whole to each team. The author supplements that the stages (identification, creation, storage, transference, and use of knowledge) are interconnected and are iterative, from the perspective that knowledge is continuously formed and altered. The author finds in the combination between KMP and DM great potential in the exploration and in the management of valuable knowledge of big data. DM provides support to KMP in the generation of inestimable knowledge and KMP provides support to DM in the collection and storage of knowledge as entrance for DM. Nonetheless, Nguyen emphasizes that there is a large gap in research in this area.

STATA (1980) discusses the concept of OL - Organizational Learning, which occurs by the means of shared insights, knowledge and mental models and that it is based on knowledge and on past experience, that is, in memory. The author elaborates that organizational memory depends on institutional mechanisms (for example, policies and strategies) used to retain knowledge, and it cannot depend exclusively on the memory of individuals, because there is always the risk of losing lessons and experiences laboriously achieved as people migrate from one job to another. Besides other motives for the attrition of personnel (retirement, separation, transfers, demise, etc.), we can also add the risk of forgetting.

DINGSØYR et al (2001) recommend that in the end of a project, an Experience Report is elaborated to collect what has gone right and what has gone wrong in the adopted process. The Review Project activity from the CRISP-DM methodology produces a similar report that, according to the description of CHAPMAN et al (2000): Resumes the important experience acquired during the project. For example, entrapment, deceitful approaches, or information to target data mining techniques more adequate to similar situations could be a part of this documentation.

According to BECKER and GHEDINI (2005), with the documentation of the DM process, as knowledge becomes explicit and managed, it increases the intellect of the organization, developing into the basis for communication and learning, supporting the dissemination of knowledge and the experiences inside the organization in different levels. The authors reinforce that the idea of capturing and storing all informal relevant knowledge generated and used during a DM process, in a way that it becomes available for subsequent recovery, constitutes an interesting approach to deal with the difficulty mentioned about reflecting in documentation the iteration and interactivity of the process.

It is a fact that companies need to educate a larger number of people about the processes and the best practices associated to DM (MARISCAL et al, 2010).

According to NGUYEN (2018), knowledge, which is an expensive commodity for organizations, with different origins, such as documents, processes, persons, communication, culture, and learning, can be incorporated in individuals and in organizations throughout processes and practices. The author states that the transference of knowledge stimulates innovation, and that the storage of knowledge is the path for the creation of an invaluable property for the organizations. An asset that is accumulated over time and that cannot be bought with any amount of money.

In view of what has been presented in the theoretical framework, it can be concluded that the documentation of the memory of a DM project is a small step, since it is located in a smaller context of a project, which can signify a giant step towards organizational memory and profits deriving from shared learning and knowledge managed for an organization. In the next section, the Rastro-DM methodology will be presented as a potential path to achieve this giant organizational step.

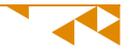

# 3.    RASTRO-DM

## 3.1    GENERAL VIEW

Rastro-DM (Rastro is a translation of trail in Portuguese) is a methodology that aims to document DM projects focused on the process of construction of models, for the purpose of rendering a trail of the executed actions, of the training performed, and of the results obtained and lessons learned. It includes three activities that correspond to the documented concepts:

• Definition of actions;
• Registration of training;
• Synthesis of learning.

Differing from traditional focus, which documents the final product (and its artifacts), the Rastro-DM methodology focuses on the documentation of the process behind the construction of models. Therefore, according to CONKLIN (1996), it promotes an increase in organizational memory, with the registration of the context of the creation of the artifacts: the objectives, the values, the experiences, the motive, the conversations, and the decisions conducted.

As observed in the theoretical framework, all the informal relevant knowledge must be documented to reflect in the documentation the iteration and interactivity of the process and be available for subsequent recovery.

The activities of the Rastro-DM methodology are complementary to the tasks projected in the methodology in use in the organization, independently of which one is used, herein called the baseline methodology, responsible for drawing all the methodological and paradigmatic frameworks used in a DM process. MINGERS and BROCKLESBY (1997) identified different ways to combine methodologies. They state that the establishment of good practices is a way of creating a new methodology; therefore, there is no inconsistency in referring to Rastro-DM as a methodology.

For the purpose of standardization and clarity, the CRISP-DM methodology will be used in the explanations as the baseline methodology, and the steps of the baseline methodology will be referred to as CRISP-DM tasks, and the steps of the Rastro-DM methodology will be referred to as Rastro-DM activities. The Rastro-DM activities will occur several times during a project and can be associated to one or more CRISP-DM tasks.

## 3.2    INTEGRATED VIEW OF THE CONCEPTS

The concepts of trail are interconnected: the training occurs in the context of a defined action in the project and these trainings can promote learning that eventually can impact, in a virtuous circle, new ideas defined in actions. Some of the possible relationships between the main concepts of the Rastro-DM mehtodology can be visualized in Figure 3. The diagram is not intended to be complete, but only to provide better understanding.

Figure 3: Main concepts associated to the Rastro-DM methodology

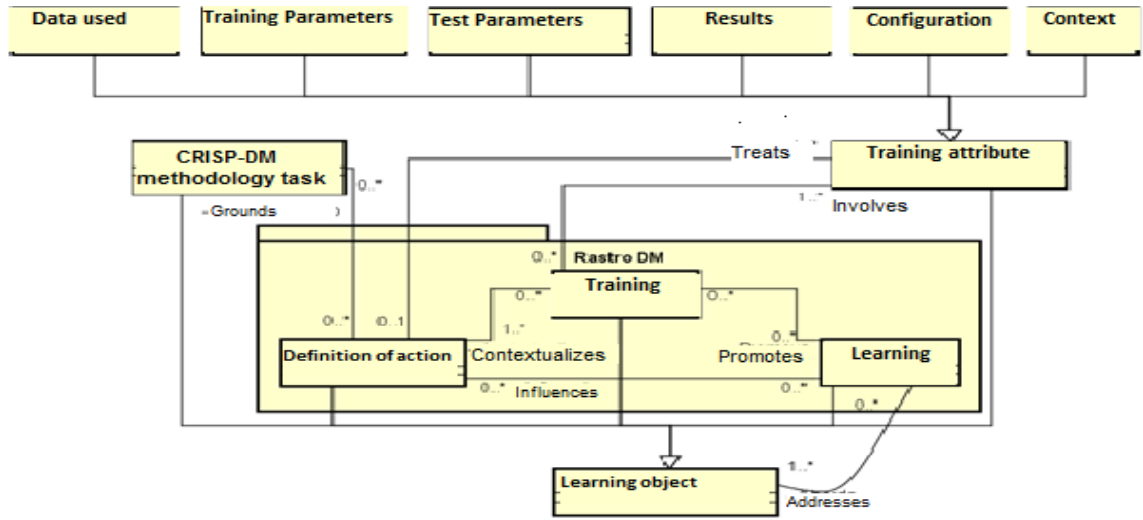

Source: Castro (2019).

It can be seen that the concept of Training attribute, subdivided into six categories, besides being related to training, can be the object for the definitions of action. The definitions of action can be founded on tasks of the baseline methodology, represented in the figure by the CRISP-DM methodology. Additionally, all the elements can be approached as learning object.

In the next section, the activities of the Rastro-DM methodology will be described.

## 3.3 DEFINITION OF ACTION

Definition of action is the activity of registration of the definition of the steps of a project, whether executed or not. The objective of the activity is not the registration of the details of the execution of an action, but the information concerning its definition, such as the statement of its objective and the techniques to be used or tested when the action is executed.

Conceptually, a definition of action corresponds to the description of one or more specific tasks of the CRISP-DM methodology instantiated within a project.

In the face of the difficulties of management in DM processes identified in the theoretical framework, the possibility of registration of the resources used in the actions defined (people, time, etc.) can be evaluated for the support of deadline estimates and the establishment of schedules of the project. Additionally, the history of these data might serve as the base for the allocation of resources in future projects.

The activity of definition of action can occur at any moment during a DM project. It is important to ensure, at a minimum, the definition of each action initiated, because, in a way, it justifies the training that will follow. Furthermore, to know what has already been carried out prevents the wastefulness of effort by repeated executions.

The level of abstraction and of detail, when more closely related to a CRISP-DM methodology task, such as, data formatting, or even more detailed, such as, formatting the name of the file for possible removal of stop words and of punctuations, will be the responsibility of each team if there is no corporate standard.

Although the simplicity of a descriptive text field is acceptable, because the content is the most important, the more structured the registration is, the better it will translate the comprehension of the process and the greater the potential of its contribution will be. The entities of the ML Schema that are reported in Figure 2 are examples of attributes associated to a Definition of Action.

To make the concepts clearer, we will use a hypothetical supervised learning project that aims to predict the price of a house (regression) for a real estate agency based in some attributes of the property.

The following examples are definitions of actions of the hypothetical project.

10/10/2018 (registration date in monty/day/year format); Experiment, as the attributes of the property in the model, the number of floors and the number of bathrooms;

11/8/2018; Evaluate if the model can achieve a better performance if the values for the distance from downtown attribute are set in a different scale;

2/1/2019; Test the lightgbm and random forest algorithms for the regression.

The comprehension of the technical terms used in examples of registrations is not relevant for the objective of this paper. The focus, in this case, is the ability of description of a specific action and not the elements of this description (lightgbm, randomforest). Nevertheless, to achieve a more complete understanding of the techniques associated to DM, the book Data Mining and Analysis: Fundamental Concepts and Algorithms, by ZAKI and MEIRA (2014) is suggested reading.

## 3.4    REGISTRATION OF TRAINING

Registration of training is the activity of documentation of the training carried out, of the parameters used and of the results obtained. In accordance with the theoretical framework, the training is the central activity for all the data mining processes, and it is important to retain the maximum of relevant information as possible from them.

The ML Schema presented in Figure 2 discloses a set of entities implicated in training and that are candidates for having their information documented in the context of the activity of registration of training.

Aiming for better comprehension, and with no intention of reaching completeness neither in classification nor in exemplification, the data involved in a training can be grouped into six categories:

- Data used: test data, training data, variables used as features for the model;

- Training Parameters: algorithms used, hyperparameters considered, implementation of techniques applied;

- Test parameters: metrics considered, forms of assessment;

- Results: model with its characteristics and the calculated metrics;

- Configuration: identification of the program, versions of the libraries used, data of hardware;

- Context: training code, date and hour, number of training periods, error message in case of interruption of the execution.

The level of detail of documentation has an impact in its utility. The code of training, a datum of the context category, can be used as identification key of the model in training.

These are examples of registrations of training of the hypothetical project for the prediction of the price of a house:

- Code: 1; Registration date: 6/7/2018; Variables used: house square footage, property square footage and Postal code of the address; Algorithm used: linear regression; Error: 0.8; Separation of test data: 10%, not stratified;

- Code: 100; Registration date: 7/7/2018; Variables used: house square footage, property square footage, Postal code of the address, number of rooms and date of the construction; Algorithm used: randomforest; Error: 0.7; Separation of test data: 5%, stratified data.

## 3.5   SYNTHESIS OF LEARNING

This is the activity of synthesis and registration of the lessons learned throughout the project, in an automatic manner or not.

The activity may occur at any moment of a DM project and can be associated or not to the training. Although they can be synthetized in the stage of pre-development of the model, in the CRISP-DM phases of business understanding and data understanding, learning, for the most part, is a result of the training runs. It is possible to outline, automatically or not, that a specific selection of variables or the use of a specified hyperparameter of a technique led to the generation of a model of better performance. There may be lessons learned that involve training that failed that aim to document how to avoid repeating errors.

Information about ML Scheme entities (Figure 2) and Defined Actions can be shown as attributes of a lesson learned.

The synthesis of learning and its effective use in the project or in future projects promote the improvement of the team in the DM process, in knowledge regarding the high level of iteration and interactivity of its tasks, regarding the different techniques and regarding the tools used.

The documentation of lessons learned prevents it to get lost in the memory of the individuals or even with the individuals when they leave the project or the organization.

To make clear the concept, we list below examples of lessons learned on the hypothetical project of price prediction of a house associated to the CRISP-DM task, as follows:

10/3/2018; Select technique (CRISP-DM task); the randomforest technique has proven itself superior to the decisiontree technique in the context evaluated;

5/26/2019; Format data; It is necessary that the values of the real estate in the data used for the training are updated to the same monetary reference;

7/26/2019; Select data; The increment of the variables relative to the number de rooms and number of parking places for automobiles promoted an increase of 10% in the accuracy of the model.

## 3.6   DIRECTIONS

In the following section, there are some practical directions for the effective and efficient application of the Rastro-DM methodology.

### 3.6.1  Adaptability

The activities of the Rastro-DM methodology are complementary to the tasks projected in the baseline methodology, which contains the entire methodological and paradigmatic framework used in a DM process. The documentation of the trail, apart from the strict context of a baseline methodology, enables a better approach of the interactivity and the iteration of the tasks, something that has not been properly mapped by DM methodologies, as discussed in the Theoretical Framework. Since the tasks and phases of a DM process merge with each other and are often executed concurrently, it is difficult to maintain an effective documentation by task of the baseline methodology. For example, if in the context of a CRISP-DM task a report is produced and subsequently we return to this task several times, the documentation would need to be constantly updated, in a way restricted to the scope of the task. The cost of the reworking makes in-time documentation impeding in the CRISP-DM methodology.

The documentation generated by the Rastro-DM methodology can connect itself to the output artifacts of the baseline methodologies. Later, the registrations can, for example, be grouped by CRISP-DM task. For example, a report named Reasons for exclusions and selections in the activity called Selection of data of the Preparation of data of CRISP-DM –phase, can be automatically generated from lessons learned and definitions of action related to data selection, supplemented by a summary of the training related to each criterion of selection experimented.

### 3.6.2  Timeliness

The registration must be in time to aid in directing the project in progress in a more efficient manner. With the adequate registration of the trail, inefficiency is avoided with repeated executions of tasks in the course of the project. As previously observed, a postmortem documentation is not adequate for DM projects, because they are considered live due to their complexity and iteration.

### 3.6.3  Flexibility

The definition of the attributes to be stored must be flexible and may vary according to the objective of mining (classification, regression, etc.), the training platform, and the objectives of the documentation. Exemplifying, if the documentation aims for the reproducibility, more details of software configuration (versions of libraries) and of random seeds used must be recorded.

Rastro-DM is flexible because it does not define a minimum proportion of attributes for each concept. After all, each project and organization has its own complexity and maturity level in DM. In a final analysis, it is the responsibility of the team to certify the veracity and the effectiveness of the documentation. It is expected that, with the improvement in DM, the organizations will be able to develop a minimum corporative standard for documentation per DM objective. However, this standard should not inhibit team creativity or merely become a burden.

### 3.6.4  Automation

It is indispensable that the activity of training registration is automatic and is connected to the development of training of model in the platform currently used. Although there may be an initial expense for the structuring of a software framework for a specific configuration of tools, this effort will be implemented only once, and the subsequent projects will benefit from it. As previously observed, GREFF et al (2017) advise on the practical challenge when the construction of such a structure is not incentivized.

It is desirable that all the activities are carried out integrated to the platforms, so that the effort is not a barrier for the documentation. After all, in order to become useful, a model of documentation must correspond, as much as possible, to the way people work (BECKER and GHEDINI, 2005).

Castro (2019) illustrates a simple implementation of an infrastructure of construction of a trail in python, created at a low cost with local files.

### 3.6.5  Usability

In reference to the definitions of action and lessons learned, their registration must be prioritized over categorization, in a way that this activity is not converted into an obstacle for their documentation. The registration can occur initially using even a free text format.

The categorization, for example, relative to the techniques used, must be carried out automatically and can be an important corporate requirement to empower the utility of the trail by means of the sharing of knowledge. As seen, the combination of DM with KMP empowers the increase of organizational knowledge (NGUYEN, 2018).

We must also consider the usability of the applications that query the trail, which tends to become corporate. Searches for information, for example, can lead to projects that have experimented a specific technique, identify information of its use, and evaluate the results attained, serving as a basis for new projects and for the definition of DM strategies in the organization.

In the next section, the application of the Rastro-DM methodology in a DM project will be illustrated.

## 4.  THE CLADOP PROJECT – ILLUSTRATION OF THE USE OF THE RASTRO-DM METHODOLOGY

The Cladop[2] project is an example of utilization of the Rastro-DM methodology. The term Cladop is used both to indicate the process (project) and to refer to the product developed (Classifier of PDF Documents). The project consisted of the development by supervised learning of an automatic classifier for PDF (Portable Document Format) documents inserted in the system for the management of Auditing of Special Accounts (e-TCE) of the Federal Court of Accounts of Brazil (TCU).

The process of Auditing of Special Accounts (TCE) ultimately seeks to reimburse the Public Treasury of damages generated by public agents, holding them accountable for these damages. With the objective of making the process of damage assessment faster and more efficient, the e-TCE system was developed by the Federal Court of Accounts of Brazil in cooperation with the Comptroller General of Brazil (CGU) through the standardization and optimization of procedures. It is a unique platform that provides access to all the Public Administration departments that act in some phase of the process of Auditing of Special Accounts. Castro (2019) details the business context and the related regulations.

The classifier developed provided benefits to businesses regarding both the usability of the system, by automatically identifying the types of documents for users, and of the data quality, by assuring greater correction of the types and higher quality of text from the documents generated by OCR (Optical Character Recognition). Documents with a correct classification of their type and with text content of higher quality are fundamental for the procedural operations assisted by machine.

---

2   After finalizing the Cladop Project for the e-TCE system, which is the presented in this paper as the "Project", the construction of a similar classifier was initiated for another system: Electronic Protocol. The new project, still in development, was named Cladop@Protocolo. Therefore, the Cladop Project presented in this paper shall be known as Cladop@eTCE.

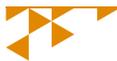

## 4.1   UNDERSTANDING THE DATA

During the construction of the Classifier, 118,266 documents were considered, relative to 4,384 damages that were distributed into 84 types of document. The type "Others" was the most used, corresponding to 18.81% of the documents. Graphic 1 presents a partial visualization of the severe unbalance of the quantity of documents between active types.

Graphic 1: Partial view of the imbalance of documents per type (reference: 4/17/2019).

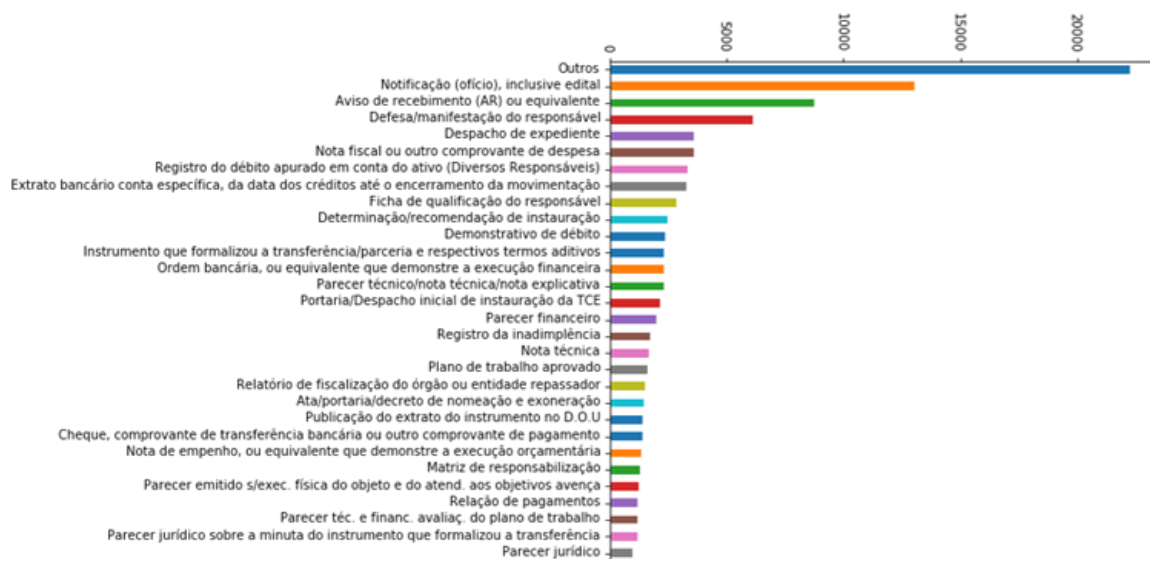

Source: Castro (2019) – partial view.

BRANTING (2017) states that the PDF format has been used in courts and that the text obtained from these documents present several errors and do not preserve the original sequence of the document, due to the OCR process used.

The outreach of the results expected by the e-TCE system depends on the quality of the documents coded in the system. The efficiency of the classifier also tends to be superior if the data have greater quality. Low quality was found in the OCR of the documents. An example of this situation is protocol document number 58.900.414 that has 162 pages, but via OCR it is impossible to identify even one hundred valid words in it, corresponding to less than one valid word per page. The low quality of the OCR can be illustrated on Graphic 2 that presents the number of valid words per page for some documents.

Graphic 2: Documents – quantity of valid words per page (y-axis) versus number of pages (x-axis).

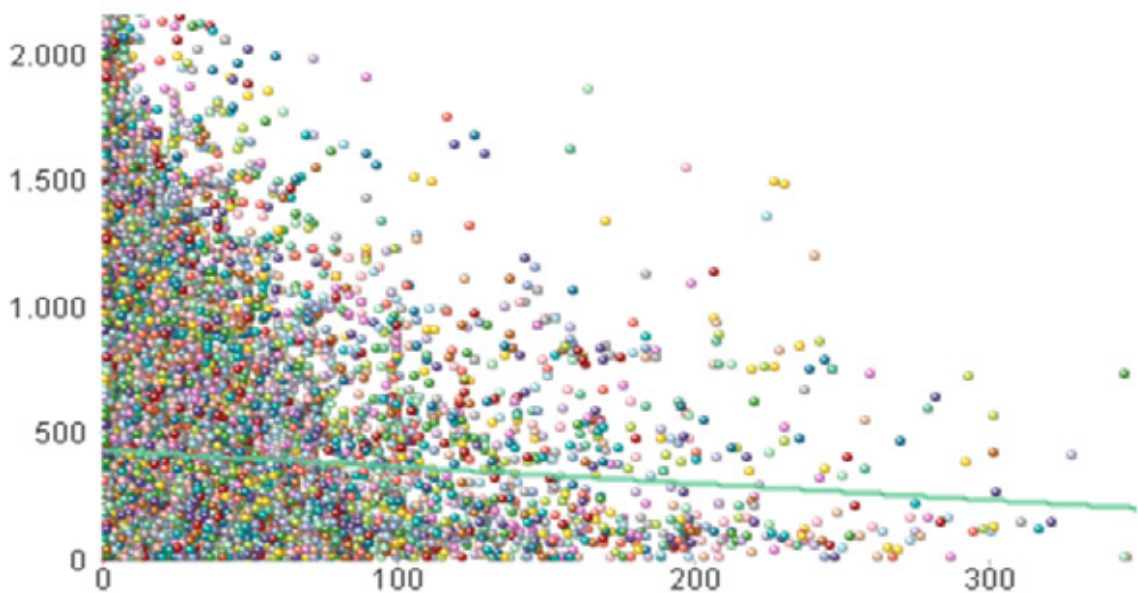

Source: the authors (2020).

According to Castro (2019), the pre-processing of the text contained in the documents was tested in different combinations of methods for obtaining text from the document (OCR algorithms), of criteria for the definition of valid words, and of substitute classes of words considered (Brazilian individual taxpayer registry identification - cpf, date, name of natural person, etc.).

## 4.2    FUNCTIONAL DESCRIPTION OF THE CLASSIFIER

The Cladop project, in its version 2.1, has achieved a level of accuracy of 91.1% with a standard deviation of 0.3%, with cross-validation of 7 partitions and 2 repetitions, totalizing 14 samples. In the context of this paper, the term accuracy must be understood as micro-accuracy, which takes into consideration the results, scores and errors, per document independently from its type.

The classifier was implemented in python in the form of a web service that receives a PDF file as its parameter, and based on its name and on its content, it returns the nine most probable types, each with their respective probabilities. The accuracy of the classifier[3] increases from 91% (for the first type) to 99% when taking into consideration the first nine predictions. The return of more than one type makes it possible for the system, for example, to present the other types in a second screen in case the first type is not accepted by the user.

Table 1 presents, for some types of documents, some metrics verified based on the validation data, 5% of the total, when the version in production of the classifier was generated. The general numbers in validation data are: accuracy: 90.99%; macro-precision: 81.04%, and weighted-precision: 91.06%; macro-recall: 80.29%, and weighted-recall: 90.99%; F1-macro: 80.16% and F1-weighted: 90.87%.

---

3    After writing this paper, there was the inclusion and exclusion of document types and, taking into consideration more recent documents, version 3.0 of the Cladop Project was developed, achieving a 93.6% accuracy level with a 0.3% standard deviation.

Table 1: Metrics verified per type regarding validation data (partial view).

| Description | Precision | Recall | F1 | Documents |
|---|---|---|---|---|
| Lawsuit - initial petition | 93.33% | 93.33% | 93.33% | 15 |
| Final ruling of higher court | 85.71% | 85.71% | 85.71% | 7 |
| Analysis of accountability | 77.27% | 62.96% | 69.39% | 27 |
| Analysis of defense | 83.33% | 52.63% | 64.52% | 19 |
| Suspension of delinquency rate | 73.68% | 82.35% | 77.78% | 17 |
| Term of concession and acceptance of scholarship and attachments | 75.00% | 54.55% | 63.16% | 11 |
| Term of definite receipt of construction work | 100.00% | 100.00% | 100.00% | 4 |

Source: Castro (2019) – partial view.

In addition to the predictions, the classifier returns information derived from the pre-processing of the file's text, which may be useful so that the e-TCE system can demand a minimum quality of text content in the documents at their registry, such as: quantity of valid words, quantity of values, quantity of names, etc. Therefore, the system can prevent low OCR quality with a high percentage of invalid words or even incomplete content in documents that are critical for the TCE process, such as the case of an Acknowledgement of Receipt document that has no date and no taxpayer registry identification.

## 4.3    APPLICATION OF THE RASTRO-DM METHODOLOGY IN THE PROJECT

The utilization of the Rastro-DM methodology has shown to be an effective solution for the difficulty of reflecting in the documentation the iteractivity and the interactivity of a DM process as presented in the theoretical framework.

The following sections exemplify the use of the methodology in the project and the benefits obtained. The details of the trail in the Cladop project (Rastro-DM@Cladop) and the code used for its construction are available at https://gitlab.com/MarcusBorela/rastro-dm.git.

### 4.3.1   Definition of action

The first definitions of action were registered as comments in the implementation code of the training. Nevertheless, throughout time, they became more complex and broader, and they crossed over different codes. Therefore, they had to remain in a databank. Definitely, the registration of definitions of action cannot rely upon coding. Due to the unpredictable nature of a DM project, the code often evolves rapidly and ends up compromising, among other things, its documentation (GREFF et al, 2017). Table 2 illustrates some definitions of action registered throughout the project.

Table 2: Examples of definitions of action registered in the Cladop project.

| Registration date | Description | CRISP-DM Task |
|---|---|---|
| 3/12/2019-17:04 | Creating a structure (code and data) for processing k-fold in shallow algorithms | Project of tests |
| 3/14/2019-19:27 | Initiated the executions to experiment optimizers (MLP): nadam, adadelta | Construct Model |
| 3/18/2019-11:40 | Experimenting with MLP columns no longer binary | Format Data |
| 3/26/2019-11:57 | Initiating the inclusion of names of files in the model | Select Data |
| 5/29/2019-19:30 | Program modified (shallow) to also record recall and f1-micro | Project of tests |
| 6/4/2019-19:35 | Program modified (shallow) to not record recall and f1-micro, because they are equivalent to accuracy (and to precision) | Project of tests |

Source: Castro (2019) – partial view.

### 4.3.2 Registration of training

The dataset structure in the databank relative to training evolved throughout the project along with the maturing of the team in DM. Figure 4 presents the main stored concepts that can be related to the subtypes of Figure 3. The concepts that are the entrance for the training are represented in light blue and the data derived from the process, in yellow.

Figure 4 – Training information registered in the Cladop project, grouped by concept

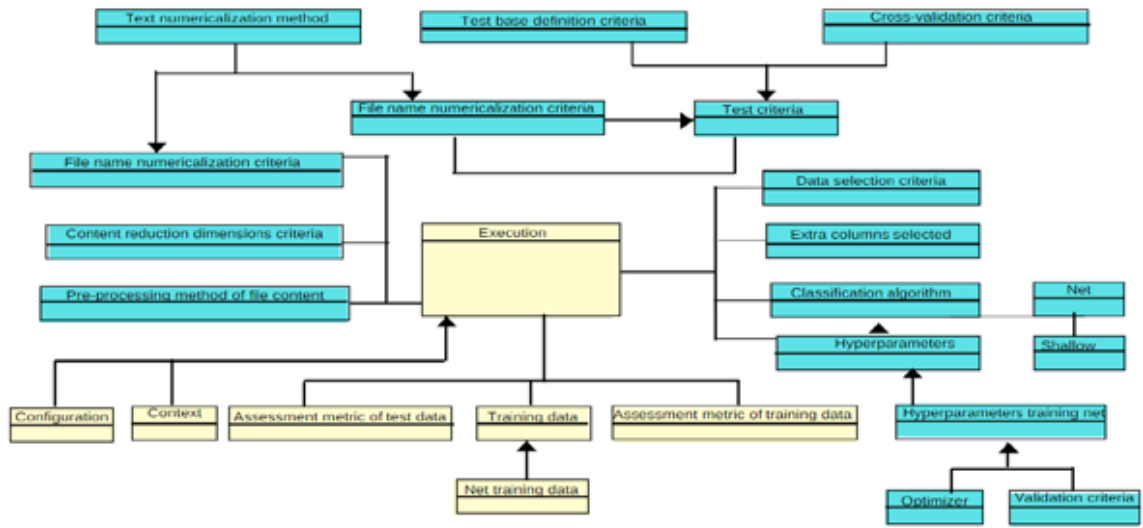

Source: Castro (2019).

Table 3 is merely illustrative and presents some attributes of these entities and their respective values for the training that tested and created the model deployed in version 2.1 of the Cladop project.

The training that generated version 2.1 of the model had as its code 13.138 (in accordance with the second line of Table 3), which demonstrates that more than thirteen thousand model training activities were executed. Some training activities involving only high-quality OCR documents and types with great quantity of documents led to classifiers with accuracy superior to 98% on test data. This is the case of training code 147, on 11/24/2018, which included 11.110 documents of better quality of OCR from the four most frequent types of documents. However, during the project, we opted for the development of a classifier that would include all the types of documents with less restrictive criteria.

Table 3 – Trail of the training for testing and for the creation of the model in version 2.1 of the Cladop project.

| Group | Item | Test | Generation |
|---|---|---|---|
| Context | Code (unique key) | 13.134 | 13.138 |
| | Date and hour of registry | 07/05/2019-00:44:59 | 07/05/2019-18:02:12 |
| | Number of periods trained | 33 | 24 |
| | Quantity of documents used | 63.468 | |
| | Quantity of types processed | 81 | |
| | Time of execution (in seconds) | 460 | 475 |
| Configuration | Program executed | Cladop_monitoramento.ipynb | |
| | Mode of text conversion for numbers | Tfidf | |
| | Number of words of the "bag of words" model of the text | 24.576 | |
| | Number of words of the "bag of words" model of the file name | 1.000 | |
| | Method of text pre-processing | 7 | |
| | Criteria for data selection | Documents of a different type from Others and created after 5/1/2018 | |
| Parameters | Method of test | Cross-validation with 7 partitions and 2 repetitions | There was no data reserved for testing |
| | Optimizer used for training the network | Adam | |
| | Size of the training batch | 256 | |
| | Minimum quantity required by type | 0 | |
| | Final dimension of the content vector after reduction | 768 | |
| | Algorithm for size reduction | TruncatedSVD | |
| | Were the data shuffled in every epoch? | No | |
| | Criteria for validation data | 5% of the data, without stratifying and with shuffle | |
| Result | Accuracy verified in test data | 91.1% +- 0.3% | |
| | Accuracy verified in training data | 96.4% +- 0.7% | 95.6% |
| | Accuracy verified in validation data | 90.8% +- 0.6% | 92.1% |

Source: Castro (2019) – adapted version.

Based on the training data, several analysis can be achieved. An example is the comparison of the performance of the models relative to the algorithms used. The models with higher accuracy in the final selection of data and parameters were implemented with neural networks (MLP network). Models implemented using the LGBMClassifier algorithm scored just behind, with a difference of approximately 2%. Graphic 3 illustrates the accuracy in test data achieved by some of the algorithms experimented. The large variation in accuracy by algorithm is explained by the variation of combination of parameters experimented in training. The understanding of these algorithms is not in the scope of this paper. To those interested in these data, we suggest the reading of the book Data Mining and Analysis: Fundamental Concepts and Algorithms from ZAKI and MEIRA (2014).

Graphic 3 - Boxplot with accuracy in test data of the algorithms experimented.

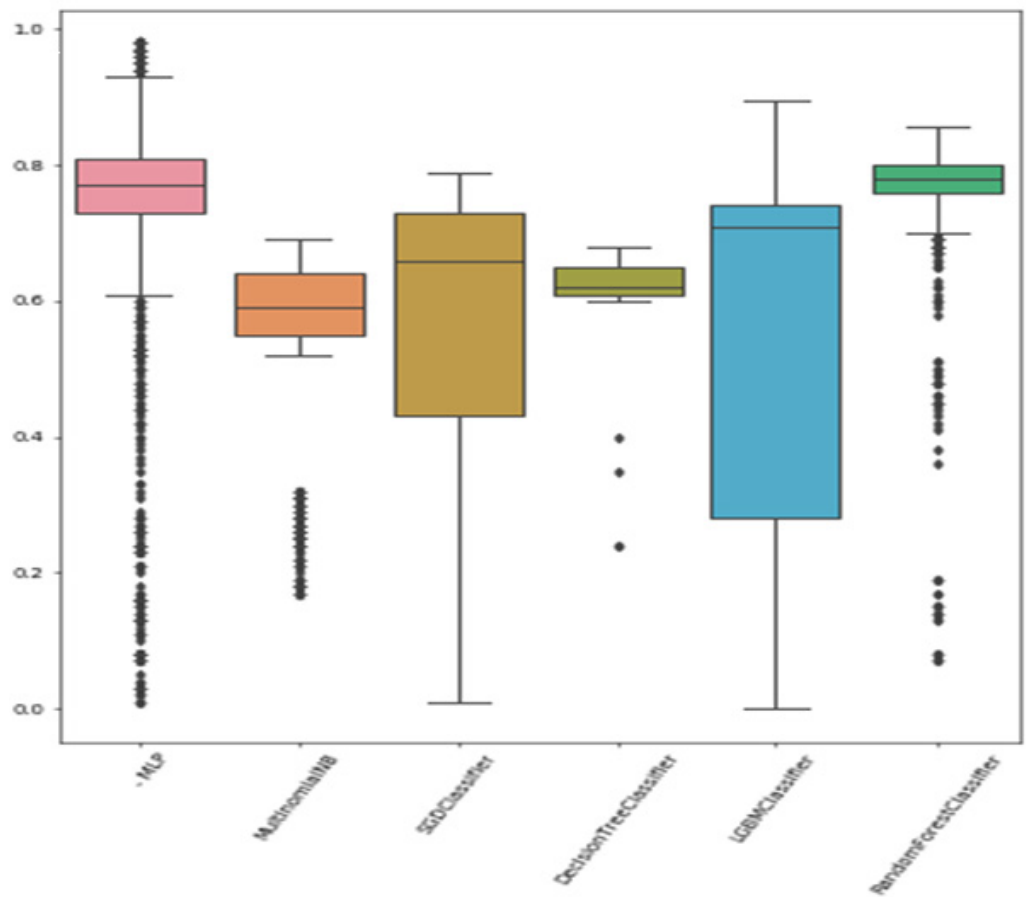

Source: Castro (2019) – partial view.

### 4.3.3 Synthesis of learning

From the lessons learned registered, the majority could be synthesized by data mining in the training dataset. For example, based on equal value queries in some columns, the leaning dated 4/10/2019 in Table 4 could be detected. Others depend on human intervention, such as the one dated 2/27/2019. The automatic generation of knowledge reveals itself to be a promising path for the improvement of organizational knowledge, which was already attested by NGUYEN (2018), that, as previously observed, emphasizes that there is a large gap in research in this area.

After less than six months after the end of the project, Castro (2019) attempted to count, by sampling, how many of these still remained complete in his memory, approaching an estimate of only 27%. Unfortunately, the remainder were either incomplete or had been lost from the author's memory. However, because of the trail, they were not lost from the memory of the project.

BECKER and GHEDINI (2005) observed that forgetting can lead to the repeated execution of experiments. Throughout the project, there was forgetting that led to unnecessary efforts in the execution of tasks. The lesson dated 4/10/2019 was ignored leading to unnecessary execution of the action to also record the metrics dated 5/29/2019 (see Table 2), which caused rework with the alteration of the program in order to not record these metrics on 4/6/2019. A lesson learned was that it is not enough to have a trail, it is also necessary to use it.

Table 4 – Examples of lessons learned registered during the project

| Registration date | Description | Context CRISP-DM activity |
|---|---|---|
| 2/27/2019-10:32 | The variables of context and PDF quality are enough to achieve a result of 44% accuracy, with optimizer Adam. With Adagrad:42%; SGD:28%. | Select Data |
| 2/27/2019-10:39 | By using pca (sklearn): it is best to separate the fit command from the transform command. The fit_transform command was locking! | Format Data |
| 2/27/2019-10:53 | By using pca (sklearn): if the number of dimensions is very small and the array is wide (see details in the documentation of function), it is better to use the randomized method rather than the full method, because it is faster and the results (variance achieved) are equivalent. | Format Data |
| 03/29/2019-09:54 | An improvement of approximately 5% in the accuracy of the models was noticed after including the file name as an attribute. | Select Data |
| 4/1/2019-15:36 | For shallow algorithms, the extra columns with no dummy values (one only column with various discrete values) led to a better result. For a Neural Networks, there is a slight improvement using dummy values. | Format Data |
| 4/10/2019-10:49 | In multiclass classification, the metrics referring to f1_micro, recall_micro, precision_micro and accuracy are equivalent. | Project of tests |
| 5/22/2019-20:04 | The Adam optimizer (passing object with Amsgrad=False) was the best optimizer so far. Better than the Amsgrad=True in 0.1%, and than the Rmsprop and the Adadelta in 0.2% in the context evaluated in the last executions. | Construct Model |

Source: Castro (2019) – partial view.

## 5.   BENEFITS ATTAINED

The benefits using the methodology in the project can be grouped into 3 aspects: project management, technical maturation of the team, and automation of activities.

Within project management, we could perceive improvement in communication, in planning, and in the reduction of expenses.

The trail made possible a continuous managerial follow-up of the project's evolution and of the performance of the models generated. The use of standardization in the submission of follow-up reports inspired reliability in users and sponsors.

The trail has become a foundation for better planning of future similar projects; after all, data mining projects can benefit from the experiences of previous projects (CHAPMAN et al, 2000). The narration of the experience contained in the trail of processes makes it possible to advocate for more realistic resources, because there is already a broader understanding of how effort is actually spent. Additionally, it enables the identification of critical points in the process.

A decrease in effort for repeated work was observed, even through unproductive paths.

The benefits of the use of the Rastro-DM methodology in management led to its inclusion as a technical requirement for outsourcing development of DM projects that has been initiated in the TCU.

The technical maturity of the DM team was increased due to the environment created, the improvements achieved, and the sharing of learning.

A stimulating and practical environment was promoted (to learn by practice). The team felt encouraged to carry out documentation in parallel with the execution of activities due to the facility in registration and the benefits perceived throughout the project, such as documentation itself.

The synthesis of learning and the effective use of these activities in the project as well as the structural evolution of the registrations of training have improved the maturity level of the team in DM. As a better comprehension and understanding of the process and of the techniques included was achieved, new fields were created for storage and the contents became more valuable.

Regarding sharing, the Cladop trail was published in the context of a Data Analysis Specialization course (Castro, 2019) and the methodology was presented in 2019 at the 5th International Conference on Data Analysis for Public Administration.

We expect that this discussion advances within the organization, relative to policies and standards to transform the trail into a corporate resource to support the dissemination of knowledge on various levels. Simple enquiries, for example, could meet projects that have applied a specific technique, and identify information from its use and assessment of the results achieved. Managerial enquiries could support DM strategies for the organization.

Another benefit was the automation of routines, which ensures a minimum quality of activities as well as reduces the cost of its realization. The chronological list of lessons learned, and definitions of action supplemented by some graphics automatically generated from the training can form automatic reports relative to tasks of the baseline methodology, as illustrated in Castro (2019). The same author demonstrates that the trail made it possible to automatically monitor the performance of the models against new documents registered in the system. Even more, this monitoring routine has promoted the automatic generation of version 2.1 of the Classifier based on more updated data.

## 6. CONCLUSION

The objectives proposed for this work were achieved. A brief contextualization by the theoretical framework was presented regarding the methodologies and documentation in data mining projects, as well as the potential impact of the experiences acquired by an organization during the projects.

Rastro-DM was proposed as a methodology for the documentation of DM projects with a focus on the process that, among other characteristics, is flexible and can be merged into the methodology used by an organization.

The Rastro-DM methodology has revealed itself viable, with beneficial results according to what was illustrated in its application in the Cladop project.

There is still a lot to be achieved. Future works may experiment the Rastro-DM methodology in other contexts: other objectives of mining rather than classification, other development platforms, or even other organizational cultures. The methodology can be incrementally reviewed with new activities or even improved by techniques and procedures that supplement its activities. Studies can be carried out for the mapping of policies and standards for trails to be used as a support for contracting DM projects. The mining of learning based on training data is also a challenge for future works because there is a large gap in research about the integration between DM and KMP (NGUYEN, 2018). Additionally, maybe the most important point for evolution, which was not the scope of this work, is how to transform a trail into a corporate resource that allows the sharing of experience and the use of institutional approaches for the development of organizational knowledge. After all, the value of the data remains in how they are interpreted and used (BERMAN et al, 2018).

Finally, it is evident that, besides contributing to the project's team, project trail has the potential of promoting a quantum leap in organizational knowledge, if institutional measures are adopted towards incentivizing its creation, its sharing, and the automatic searching of the learning contained in them. Nonetheless, we must be mindful that no corporate strategy should constrain the liberty of the data analysts, because it diminishes their creativity. And the creative freedom for developing their own trails is something that humans cannot do without, since it is one of the traits that prevents us from being classified as machines.

## REFERENCES

BECKER, Karin; GHEDINI, Cinara. **A documentation infrastructure for the management of data mining projects**. Information and Software Technology, v. 47, n. 2, p. 95-111, 2005.

BERMAN, Francine et al. **Realizing the potential of data science**. Communications of the ACM, v. 61, n. 4, p. 67-72, 2018.

BHATT, Ganesh D. Knowledge management in organizations: examining the interaction between technologies, techniques, and people. Journal of knowledge management, v. 5, n. 1, p. 68-75, 2001.

BRANTING, K. **Automating Judicial Document Analysis**. ASAIL@ICAIL. 2017.

CASTRO, Marcus. **Mineração de Dados com Rastro: Boas Práticas para Documentação de Processos e sua Aplicação em um Projeto de Classificação Textual.** Conclusion paper of the specialization program on Data Analysis for Auditing. Brazilian Federal Court of Account's Serzedello Corrêa Institute, Brasília, Brazil. Available at: https://portal.tcu.gov.br/biblioteca-digital/mineracao-de-dados-com-rastro-boas-praticas-para-documentacao-de-processo-e-sua-aplicacao-em-um-projeto-de-classificacao-textual.htm. Accessed on July 26, 2020. 2019.

CHAPMAN, Pete *et al*. **CRISP-DM 1.0: Step-by-step data mining guide**. SPSS inc, v. 16, 2000.

CHOLLET, Francois. **Deep Learning with Python**. Manning Publications Co., Greenwich, CT, USA. 2017.

CONKLIN, Jeffret. **Capturing Organizational Memory**. In: Groupware and Computer-Supported Cooperative Work, R.M. Barcker (Ed.), Morgan Kaufman, pp. 561-565. 1996.

DINGSØYR, Torgeir; Moe, Nils Brede; Øystein. Nytrø. **Augmenting experience reports with lightweight postmortem reviews**. Lecture Notes in Computer Science, 2188:167–181, 2001.

GHEDINI, Cinara; BECKER, Karin. **KDD application management through documentation**. Available at: https://www.researchgate.net/profile/Karin_Becker2/publication/268253354_KDD_application_management_through_documentation/links/5657a5ec08ae1ef9297bf1d1/KDD-application-management-through-documentation.pdf. Accessed on July 27, 2019. 2000.

\_\_\_\_\_\_. **A documentation model for KDD application management support**. In: SCCC 2001. 21st International Conference of the Chilean Computer Science Society. IEEE. p. 105-114. 2001.

GREFF, Klaus *et al*. **The sacred infrastructure for computational research**. In: Proceedings of the Python in Science Conferences-SciPy Conferences. 2017.

KURGAN, Lukasz A.; MUSILEK, Petr. **A survey of Knowledge Discovery and Data Mining process models**. The Knowledge Engineering Review, v. 21, n. 1, p. 1-24, 2006.

MARBÁN, Óscar *et al*. **An engineering approach to data mining projects**. In: International Conference on Intelligent Data Engineering and Automated Learning. Springer, Berlin, Heidelberg, p. 578-588. 2007.

MARISCAL, Gonzalo; MARBAN, Oscar; FERNANDEZ, Covadonga. **A survey of data mining and knowledge discovery process models and methodologies**. The Knowledge Engineering Review, v. 25, n. 2, p. 137-166, 2010.

MINGERS, John; BROCKLESBY, John. **Multimethodology: Towards a framework for mixing methodologies**. Omega, v. 25, n. 5, p. 489-509, 1997.

NGUYEN, Ngoc Buu Cat. **Data Mining in Knowledge Management Processes:** Developing an Implementing Framework, 2018.

PRAKASH, BV Ajay; ASHOKA, D. V.; ARADHYA, VN Manjunath. **Application of data mining techniques for software reuse process**. Procedia Technology, v. 4, p. 384-389, 2012.

PUBLIO, Gustavo Correa et al. **ML-Schema: Exposing the Semantics of Machine Learning with Schemas and Ontologies**. arXiv preprint arXiv:1807.05351, 2018.

STATA, Ray. **Organizational learning: The key to management innovation**. Massachusetts Institute of Technology, 1980.

W3C (World Wide Web Consortium) Machine Learning Schema Community Group. **W3c machine learning schema**. Available at: https://www.w3.org/community/ml-schema. Accessed on July 30, 2019. 2017.

WIRTH, Rüdiger; HIPP, Jochen. **CRISP-DM: Towards a standard process model for data mining**. In: Proceedings of the 4th international conference on the practical applications of knowledge discovery and data mining. Citeseer, p. 29-39. 2000.

ZAKI, Mohammed J.; MEIRA, Wagner. **Data mining and analysis: fundamental concepts and algorithms**. Cambridge University Press, 2014.